\documentclass[conference]{IEEEtran}
\IEEEoverridecommandlockouts
\usepackage{cite}
\usepackage{amsmath,amssymb,amsfonts}
\usepackage{algorithmic}
\usepackage{graphicx}
\usepackage{textcomp}
\usepackage{xcolor}
\usepackage[utf8]{inputenc} 
\usepackage[T1]{fontenc}    
\usepackage{hyperref}       
\usepackage{url}            
\usepackage{booktabs}       
\usepackage{nicefrac}       
\usepackage{microtype}      
\usepackage{lipsum}
\usepackage{fancyhdr}       
\usepackage{graphicx}       
\usepackage{stfloats}
\usepackage{verbatim}
\usepackage{indentfirst}\usepackage{indentfirst}
\usepackage{balance}
\usepackage{subcaption}

\usepackage{fancyhdr}
\usepackage{lipsum}
\usepackage[justification=centering]{caption}
\usepackage[a4paper, total={184mm,239mm}]{geometry}
\def\BibTeX{{\rm B\kern-.05em{\sc i\kern-.025em b}\kern-.08em
    T\kern-.1667em\lower.7ex\hbox{E}\kern-.125emX}}
\begin{document}

\pagestyle{fancy}
\fancyhf{}
\fancyhead[C]{PREPRINT - Accepted for publication at the 2026 Design, Automation and Test in Europe Conference (DATE)}

\fancypagestyle{plain}{
	\fancyhf{}
	\fancyhead[C]{PREPRINT - Accepted for publication at the 2026 Design, Automation and Test in Europe Conference (DATE)}
	\fancyfoot[C]{\thepage}
}

\title{MPM-LLM4DSE: Reaching the Pareto Frontier in HLS with Multimodal Learning and LLM-Driven Exploration}

\author{
	\IEEEauthorblockN{Lei Xu, Shanshan Wang, Chenglong Xiao$^{*}$}
	\IEEEauthorblockA{Department of Computer Science, Shantou University, China}
	\{24lxu,sswang,chlxiao\}@stu.edu.cn\\
	\thanks{*Corresponding author: Chenglong Xiao (chlxiao@stu.edu.cn). This work is partially sponsored by Guangdong Basic and Applied Basic Research Foundation (2022A1515110712, 2025A1515010272 and 2023A1515010077, the Scientific Research Project of Colleges and Universities in Guangdong Province of China under Grant No. 2021ZDZX1027).}}
\maketitle
\thispagestyle{fancy}

\begin{abstract}
	High-Level Synthesis (HLS) design space exploration (DSE) seeks Pareto-optimal designs within expansive pragma configuration spaces. To accelerate HLS DSE, graph neural networks (GNNs) are commonly  employed as surrogates for HLS tools to predict quality of results (QoR) metrics, while multi-objective optimization algorithms expedite the exploration. However, GNN-based prediction methods may not fully capture the rich semantic features inherent in behavioral descriptions, and conventional multi-objective optimization algorithms often do not explicitly account for the domain-specific knowledge regarding how pragma directives influence QoR.  To address these limitations, this paper proposes the MPM-LLM4DSE framework, which incorporates a multimodal prediction model (MPM) that simultaneously fuses features from behavioral descriptions and control and data flow graphs. Furthermore, the framework employs a large language model (LLM) as an optimizer, accompanied by a tailored  prompt engineering methodology. This methodology incorporates pragma impact analysis on QoR to guide the LLM in generating high-quality configurations (LLM4DSE). Experimental results demonstrate that our multimodal predictive model significantly outperforms state-of-the-art work ProgSG by up to 10.25$\times$. Furthermore, in DSE tasks, the proposed LLM4DSE achieves an average performance gain of 39.90\% over prior methods, validating the effectiveness of our prompting methodology. Code and models are available at \url{https://github.com/wslcccc/MPM-LLM4DSE}. 
\end{abstract}


\section{Introduction}
	High-Level Synthesis (HLS) has emerged as a paradigm-shifting methodology in modern VLSI design. This approach empowers designers to utilize algorithmic specifications in high-level languages (e.g., C/C++) for hardware generation, while providing tunable synthesis parameters to optimize implementation quality. However, the substantial time required by HLS tools to generate quality-of-results (QoR) metrics remains a critical bottleneck particularly during hardware optimization where design spaces are often combinatorially vast. Within design space exploration (DSE), most algorithms struggle with both aspects due to their inability to interpret the impact of synthesis pragmas on final implementations. Thus, the pivotal objectives for HLS DSE are to rapid and accurate QoR prediction and establish an intelligent DSE paradigm.

	\begin{figure}[t]
		\centering
		\includegraphics[width=0.65\linewidth]{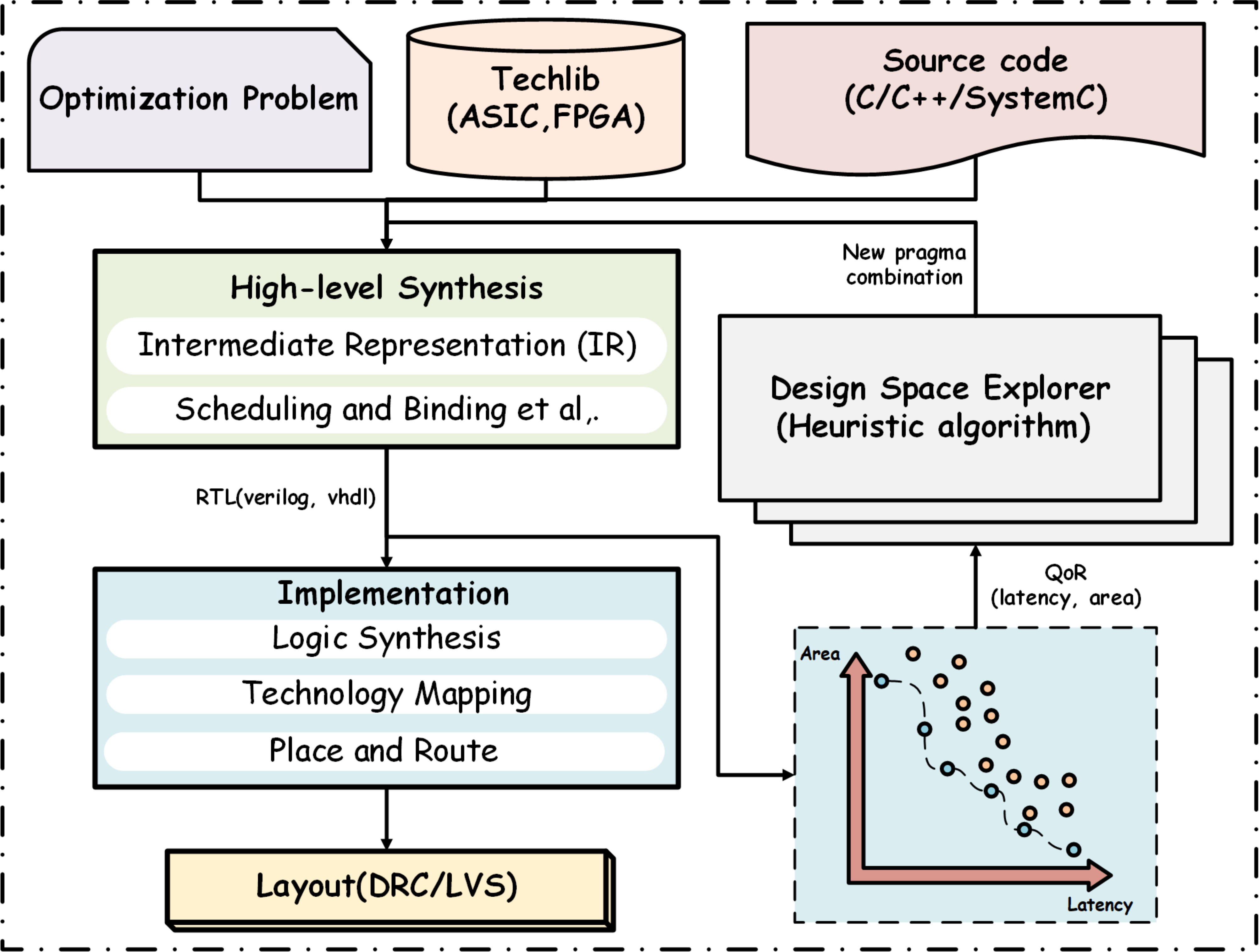}
		\caption{An overview of HLS DSE.}
		\label{HLSDSE1}
		\vspace{-6mm}
		
	\end{figure}
	As illustrated in Fig. \ref{HLSDSE1}, the HLS DSE flow begins with directive-annotated C/C++ behavioral descriptions, which undergo high-level synthesis to generate RTL implementations and extract corresponding QoR metrics. A multi-objective optimization engine then explores new directive combinations, iteratively refining the design space to converge on Pareto-optimal configurations that balance QoR metrics. Notably, deep learning techniques are employed to accelerate this process by rapidly predicting QoR metrics, thus reducing the need for full synthesis at each iteration.
	
	{\bf{QoR prediction.}} The evolution of QoR prediction methodologies demonstrates a clear trajectory:  starting from foundational graph neural network (GNN)-based feature extraction on control and data flow graphs (CDFGs), it has progressed toward increasingly sophisticated architectures and paradigms. Early works establish GNNs as viable surrogates for HLS QoR estimation (Wu et al. \cite{Wu2021}), with subsequent innovations enhancing topological awareness through edge-centric aggregation (Lin et al. \cite{Lin2022}) and hierarchical pooling (Kuang et al. \cite{Kuang2023}). Breakthrough methodologies then emerged that bypassed traditional EDA dependencies: Sohrabizadeh et al. achieved millisecond-level evaluation via surrogate modeling \cite{Sohrabizadeh2022}, Ferretti et al. matched industrial simulator accuracy without compiler internals \cite{Ferretti2022}, and Gao et al. pioneered direct C/C++ source processing with multi-granular embeddings \cite{Gao2024}. Most recently, Qin et al. proposed a representation learning method combining the source sequence and the CDFG (ProgSG \cite{Qin2024}). 
	
	{\bf{DSE methods.}} Prior DSE methodologies  have primarily built upon evolutionary algorithms and machine learning. Wang et al. utilized machine learning to optimize metaheuristic parameters, demonstrating significant performance gains over hand-tuned approaches \cite{Wang2020}. Wu et al. established an reinforcement learning driven framework for Pareto-optimal resource allocation across competing objectives \cite{Wu2023}, while Yao et al. decomposed DSE problems via MOEA/D and EDA probabilistic modeling to minimize synthesis runs \cite{yao2025}. Xu et al. introduced the LLMMH framework, which integrates large language models (LLMs) as solution operators within metaheuristics to enhance DSE precision \cite{xu2025}.

    
    While GNN-based QoR prediction has shown promise, it still has several limitations. CDFGs only provide control and data flow information to GNNs and do not explicitly capture structural semantics or programmer annotations, which are critical for prediction accuracy. For instance, with directives such as \verb|#pragma HLS UNROLL factor=4|, the CDFG contains \verb|UNROLL| nodes connected to corresponding code blocks. However, graph feature extractors struggle to capture both the precise semantic meaning and operational scope of \verb|UNROLL| directives. While \verb|UNROLL| node insertion supplements abstract CDFG graph information, it fails to accurately represent the implicit intent behind pragma usage in the source code. Although ProgSG \cite{Qin2024} introduced token-node alignment to mitigate this issue, its monolithic transformer architecture may not fully capture complex source code features and lacks robust mechanisms for fusing graph and text representations. For example, ProgSG's method merely converts the \verb|UNROLL| directive into a node and aligns it with corresponding tokens, failing to capture how the directive influences code statement blocks. Furthermore, while DSE algorithms aim to iteratively improve solutions, many existing methods often do not systematically account for the significant influence of pragma directives on final QoR metrics such as latency and resource utilization. To address the aforementioned challenges, we propose MPM-LLMDSE, an automated framework featuring:
	\begin{itemize}
		\item [1)] \textbf{A Multimodal Dataset}: Constructs Graph-Text dataset combining CDFG structural features with semantic embeddings extracted from pragma-augmented source code using a pre-trained language model (LM).
		\item [2)] \textbf{A Multimodal Prediction Model (MPM)}: Proposes a hybrid architecture leveraging LMs and GNNs to extract complementary features from source code and CDFG respectively, dynamically fused via multi-head attention mechanisms.
		\item [3)] \textbf{An LLM-Driven DSE Engine (LLM4DSE)}: Introduces LLM as an optimizer for DSE, and a sophisticated prompt engineering methodology (PEODSE) that incorporates pragma impact analysis and high-quality examples to guide LLMs in generating high-quality design configurations.
	\end{itemize}
	
\section{PRELIMINARIES}		
	\subsection{Multimodal Graph Learning In HLS QoR Prediction}\label{II-B}
    Existing GNN-based HLS QoR prediction methods, which rely on CDFGs generated by compilation tools \cite{Lattner}, \cite{Chris}, primarily utilize structural flow information but do not fully capture the rich semantic features available in source code. This issue motivates our investigation into more effective integration of source-level semantics for improved prediction accuracy. Tang et al. \cite{tang2025} demonstrate LLM embeddings' intrinsic regression capability with MLP heads, while Joshi et al. \cite{Joshi2025} conceptualize Transformers as message-passing GNNs. Transformer-based LMs (e.g., BERT \cite{Devlin2019})  can leverage their inherent multi-head attention mechanisms to learn semantic and syntactic features from text \cite{clark2019}. Building upon these theoretical foundations, we posit that fusing graph representations and textual representations yields embeddings incorporating both semantic information and graph structural information, thus leading to more accurate QoR prediction. However, graph representations and textual representations reside in distinct representation spaces, making their effective integration the critical challenge. Based on the preceding analysis, we conceptualize multimodal graph learning for QoR prediction as follows:	
    \begin{equation}
		Target_{i} = MLP_{i} \left(Fuse(h_{\mathcal{G}}, h_{\mathcal{S}})\right) 
    \end{equation}
	where $i$ denotes the index of the QoR prediction objective, $h_{\mathcal{G}}$ and $h_{\mathcal{S}}$ represent the global structural representations extracted by GNNs and global semantic representations derived from LMs respectively, and $Fuse$(·) denotes the parametric fusion mechanism integrating $h_{\mathcal{G}}$ and $h_{\mathcal{S}}$ to yield joint representations.
    \begin{figure}[t]
		\centering
		\includegraphics[width=0.95\linewidth]{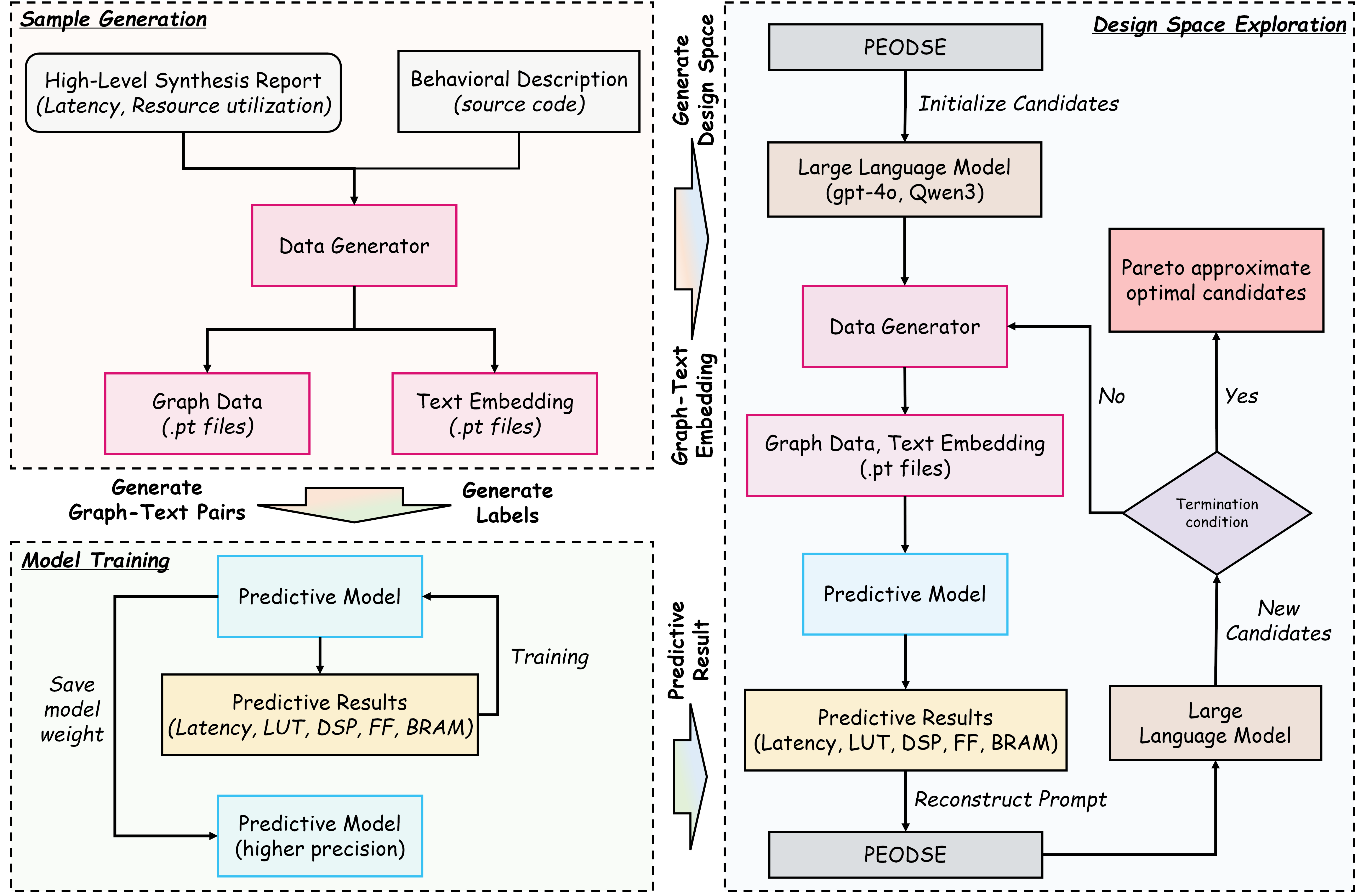}
		\caption{The framework of MPM-LLM4DSE.}
		\label{framework}
		\vspace{-6mm}
	\end{figure}
    \vspace{-6mm}
\subsection{Multimodal Dataset In HLS QoR Prediction}
    Traditional single-CDFG datasets are incompatible with our proposed predictive model, thus we introduce a novel multimodal dataset specifically designed for QoR prediction. Given a CDFG $\mathcal{G}$, its design configuration $d_{c}$, and behavioral description $b_{d}$, we first merge the configuration $d_{c}$ and description $b_{d}$ into composite text inputs. After tokenization, these inputs are passed through the LM. $h_{\mathcal{S}}$ is derived by averaging the summed hidden states of the CLS token from the final layer of the LM (CLS token inherently represents the global semantic representation of the input text). The process is formulated as:
	\begin{equation}\label{eq3}
		h_{\mathcal{S}} = \frac{1}{l} \sum_{k=0}^{l} CLS(LM_{hidden_{k}}(tokenizer\left( merge\left( d_{c}, b_{d} \right) \right) ))
	\end{equation}
    where $hidden_{k}$ is the output of the last hidden state of kth layers of the LM, $merge$(·)  is used to merge behavioral descriptions and design configurations, $tokenizer$(·) is used to convert text data into tokens and $CLS$(·) gets the CLS token features from the hidden states of a LM. Combining the obtained feature representations $h_{\mathcal{S}}$ with corresponding CDFG $\mathcal{G}$ yields individual trainable data instances.
	
\section{Methodology}	
	Fig. \ref{framework} shows the MPM-LLM4DSE framework that integrates three cohesive modules: Sample Generation employs compilation tools and GNNs to extract graph representations from sources codes, while language model is used to extract textual representations, thereby forming multimodal embedding representations. These representations are then combined with QoR metrics extracted from HLS reports to create the training samples. Model Training initializes model weights and optimizes them through training on generated multimodal datasets, persistently storing weights that minimize prediction error. Design Space Exploration orchestrates an iterative loop where task-specific prompts guide LLMs to generate design configurations until the given termination condition is met (e.g., convergence thresholds or iteration limits). Each iteration involves the Data Generator synthesizing these with source code into graph-text embedding representations, and the predictive model evaluating corresponding QoR metrics.

    \begin{figure}[t]
		\centering
		\includegraphics[width=0.40\linewidth]{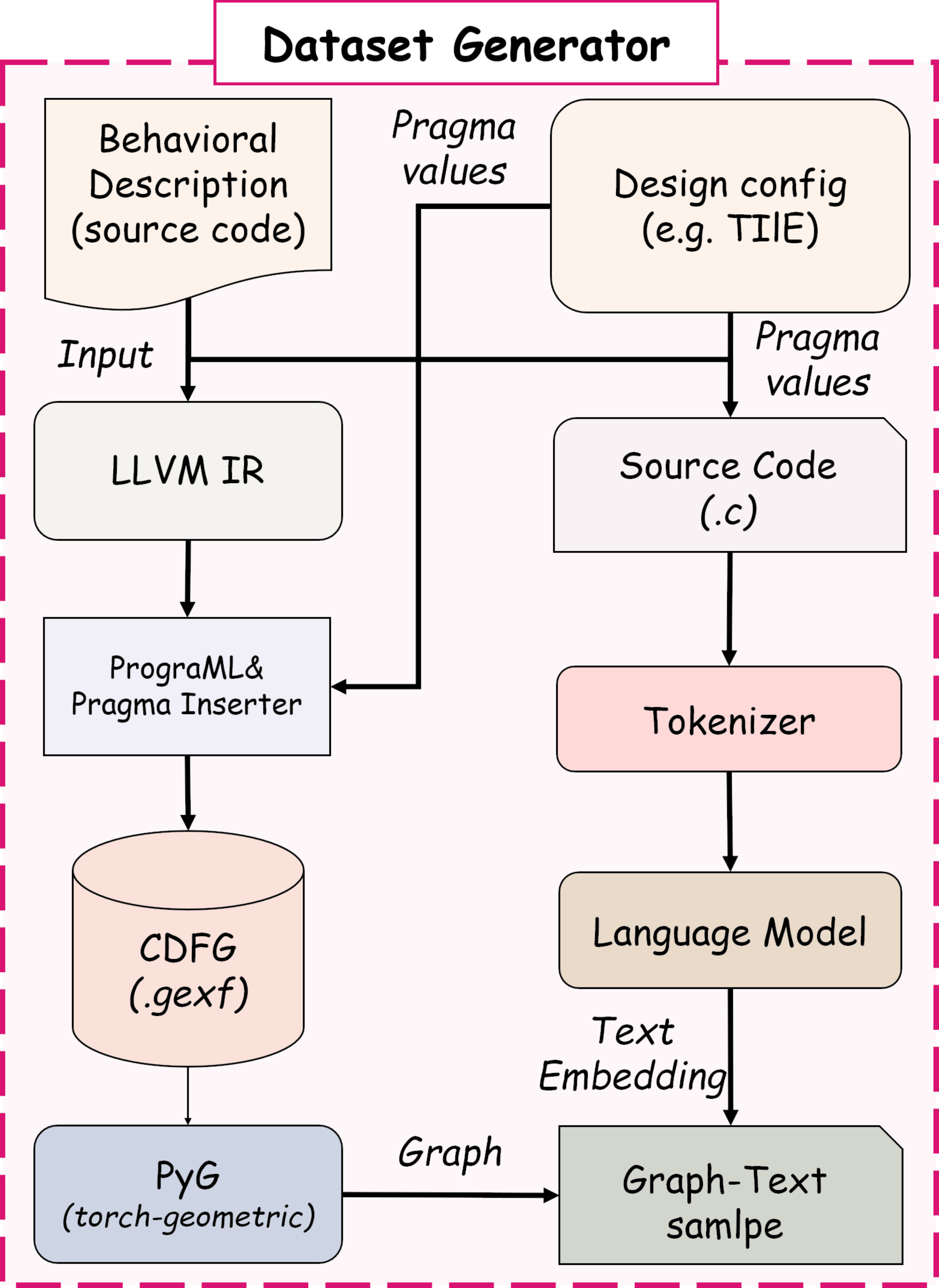}
		\caption{The workflow of Dataset Generator.}
		\label{DG}
		\vspace{-2mm}
	\end{figure}

	\begin{table}[t]
		\footnotesize
		\setlength{\tabcolsep}{0.1pt}
		\centering
		\caption{Node features of graph data.}
		\begin{tabular}{c c c}
			\toprule
			\bf{Features} & \bf{Description} & \bf{Values type} \\
			\midrule
			\textit{node type} & index of node type & long \\ 
			\textit{instruction type} & index of instruction & long \\
			\textit{function type} & index of function & long \\
			\textit{block type} & index of block & long \\
			\textit{Latency} & amount of clock cycles in the node & int \\
			\textit{LUT} & amount of LUT used in the node & int \\
			\textit{DSP} & amount of DSP in the node & int \\
			\textit{FF} & amount of FF in the node & int \\
			\bottomrule
		\end{tabular}
		\vspace{-5mm}
		\label{t1}
	\end{table}
	
	\subsection{Data Generator}\label{III-A}
		Fig. \ref{DG} outlines the processing flow of the Data Generator. The source code is first processed by LLVM to produce an intermediate representation (IR), which is then converted into a CDFG via ProGraML. Concurrently, in accordance with the GNN-DSE \cite{Sohrabizadeh2022}, pragma inserter constructs icmp nodes to enrich the CDFG with pragma-related information. On the other hand, as discussed in Equation \ref{eq3}, we obtain complete source code by integrating pragmas. 
        
        While LLMs rapidly evolve with parameter counts reaching hundreds of billions, most of LMs contain merely millions of parameters. However, for specialized downstream tasks, LMs trained on domain-specific corpora have been shown to outperform LLMs. Furthermore, owing to their faster deployment and reduced computational demands, such LMs represent a highly suitable choice for QoR prediction applications. In this work, we choose the pre-trained CodeBERT-c \cite{Zhou2023} as our language model because the behavioral descriptions are typically written in C or C++. CodeBERT-c is specifically trained on C code, exhibiting stronger comprehension of the C language. Therefore, selecting CodeBERT-c enables better feature extraction from the source code. Using a tokenizer, we convert the source code into token representations, and employ CodeBERT to extract and save features from these token representations. By integrating CDFGs and text embeddings, we obtain trainable Graph-Text representations. The Graph-Text representations along with the QoR metrics in HLS reports or predictive results are used to form the training samples. Additionally, the node features of CDFGs are presented in Table \ref{t1}. Categorical features such as node type and instruction type are represented using one-hot encoding, while preprocessed numerical features directly utilized.
	\begin{figure}[t]
		\centering
		\includegraphics[width=0.82\linewidth]{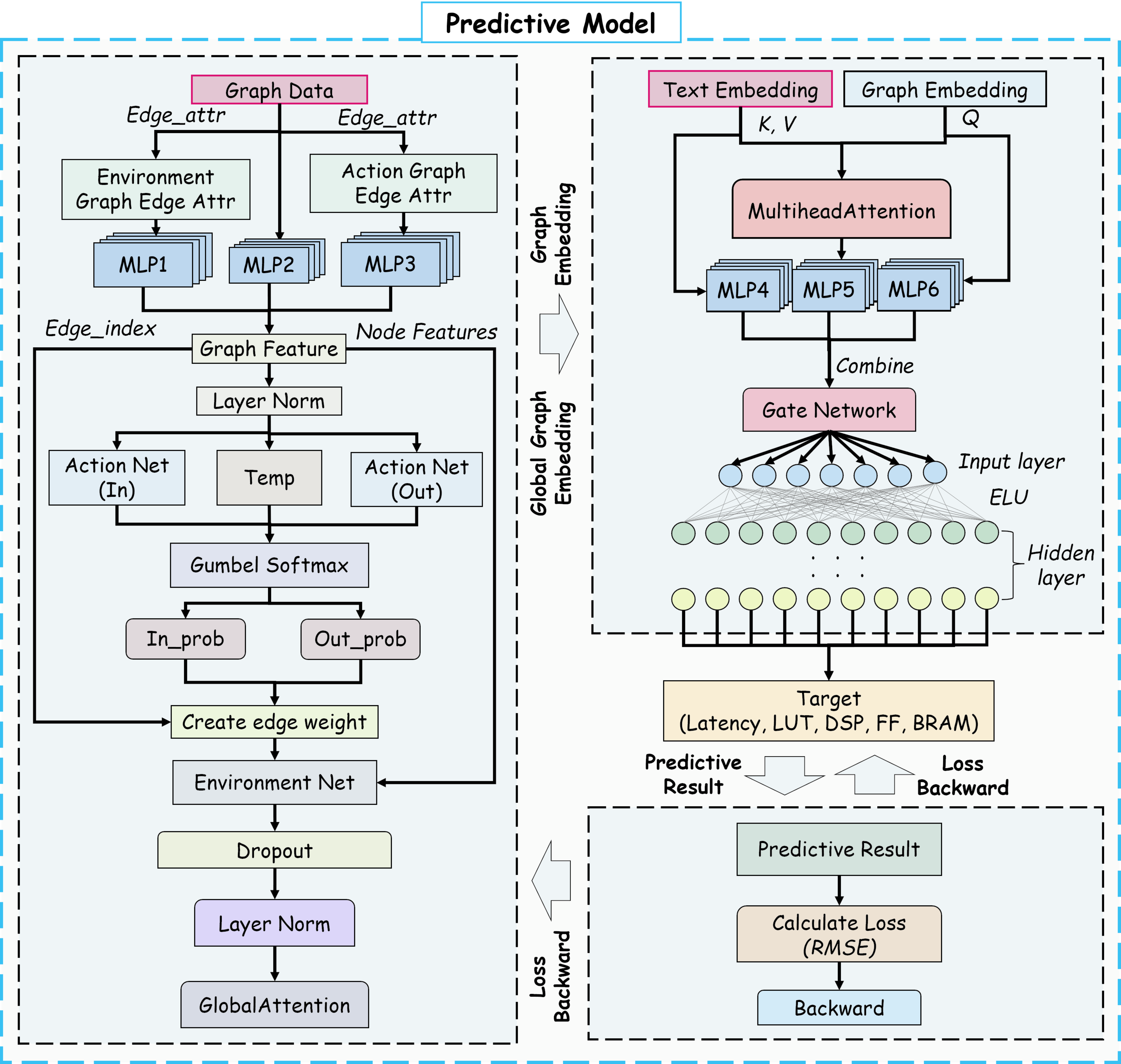}
		\caption{The architecture of the proposed predictive model.}
        \vspace{-6mm}
		\label{model}
	\end{figure}

	\subsection{Multimodal Predictive Model} \label{III-B}

        Fig. \ref{model} illustrates the architecture of the proposed multimodal predictive model, which consists of three core components: graph feature extraction using GNNs, fusion of graph features with text embeddings, and the end-to-end training of the predictive model. Firstly, prior to the GNN processing, the Graph Data undergoes preliminary transformation in which the original edge feature matrix is mapped into two distinct edge feature matrices. These, along with the original graph structure, collectively form the enhanced Graph Feature representation used in subsequent stages.
		
		Secondly, we design a Enhanced-CoGNN (ECoGNN) variant for QoR prediction inspired by the CoGNN \cite{Finkelshtein2024} framework. Since the baseline CoGNN only processes undirected graph information flow, we extended its functionality. We redefine the set of node states $\Phi$ as follows:
		\begin{equation}
			\Phi = \{S, L_{in}, L_{out}, B, I\}
		\end{equation}
		where $S$, $B$, and $I$ denote node states for information aggregation: $S$ receives neighbor information while broadcasting its own state, $B$ exclusively broadcasts its own state without receiving, $I$ neither broadcasts nor receives information; $L_{in}$ receives exclusively from neighbors connected via incoming edges, and $L_{out}$ receives exclusively from neighbors connected via outgoing edges. 
		
		At the $k$-th ECoGNN layer, we input node features $h_{k-1}$ into LayerNorm \cite{Ba2016} to obtain the regularized feature $h_{k-1}^{'}$. Since node actions follow a categorical distribution rendering the training process non-differentiable. We employ the Gumbel-Softmax estimator \cite{Jang2017} to achieve differentiable training. According to comparative experiments in \cite{xu2025}, Environment Net and Action Net are implemented using MEANGUNs \cite{Hamilton} and SUMGNNs \cite{Hamilton} respectively, with Temp (learnable Gumbel softmax temperature) primarily realized via MLPs. The weight of the edges $edge\_weights$ are constructed for outgoing and incoming edges based on the index of the edges $edge\_indexs$  and the probability of information inflow and outflow $prob_{in}$, $prob_{out}$, integrated with the $\mathcal{G}_{f}^{'}$ and fed into the Environment Net to yield $h_{i}$, which is then regularized via LayerNorm. By iterating the above process until node features are output from the final ECoGNN layer, we obtain final node-level representations $h_{f}$. To derive graph-level embeddings $h_{\mathcal{G}}$, these would directly feed into an MLP for QoR prediction. However, using average pooling causes significant information loss. To address this, we leverage global node attention \cite{Li2016} to generate the final graph-level embedding $h_{\mathcal{G}}$.
        With both textual embeddings from source code and graph-level embeddings available, we prioritize multimodal feature fusion. Direct summation introduces excessive feature noise. In image captioning, Sun et al. employed multi-head attention to align image regions with tokens \cite{bao2022}. As defined for QoR-oriented multimodal graph learning, our fusion aims to supplement semantic representations for nodes in graph data and augment structural information for subregions. Multi-head attention \cite{vaswani2023} effectively fulfills this objective by fusing graph-level and textual embeddings. 
		
		Then we propose using the graph embedding $h_{\mathcal{G}}$ as the query matrix $Q$ and the text embedding as the key matrix $K$ and value matrix $V$. Multi-head attention computation is performed and the results are aggregated to obtain fused features $h_{fuse}$. The process is formulated as:
        \begin{equation}
		head_{w} = Attention(QW_w^Q, KW_w^K, VW_w^V)
			\vspace{-5mm}
	   \end{equation}
		
	   \begin{equation}
		h_{fuse} = Concat(head_{1},...,head_{w})W^O	
	   \end{equation}
        where $W_w^Q$, $W_w^K$ and $W_w^V$ are the learnable weight matrices of  $Q$, $K$ and $V$ respectively, $W^O$ is used to fuse the outputs of multiple heads. MLPs first align the dimensions of the graph embedding $h_{\mathcal{G}}$ and the fused features $h_{fuse}$. These aligned features are then fed into a gated network \cite{Li2016} to dynamically control the information flow between them.
        
		
		Finally, the dynamically fused feature $h_{fuse}^{'}$ is fed into an MLP prediction head to yield predicted results. The root mean squared error (RMSE) loss is computed to update model parameters. While mean absolute percentage error (MAPE) also reflects prediction accuracy, for targets like latency with values exceeding hundreds of thousands, a 1\% error translates to thousands of clock cycles. Thus, RMSE more accurately quantifies prediction error.
		
\begin{figure}[t]
			\centering
			\includegraphics[width=0.6\linewidth]{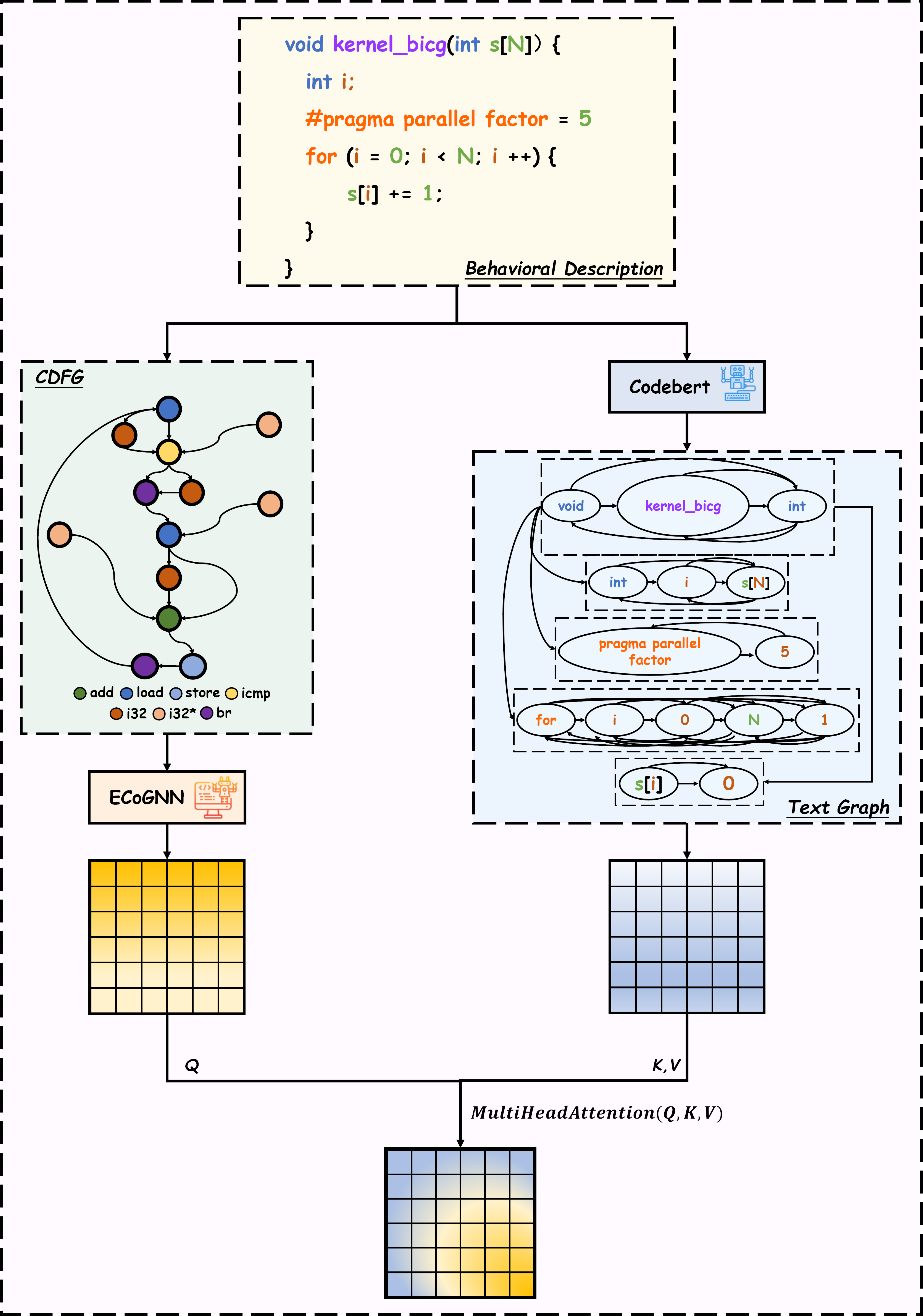}
			\caption{Multimodal feature fusion for QoR prediction.}
			\label{mi}
			\vspace{-5mm}
		\end{figure}

		Fig. \ref{mi} depicts the process of multimodal feature fusion for QoR prediction. The left portion of Fig. \ref{mi} illustrates ECoGNN's feature extraction process from CDFGs, with the resulting features forming the Query matrix. The right section demonstrates LM's feature extraction from source code, where the text graph \cite{Joshi2025} clarifies that this process inherently operates as a message-passing procedure analogous to GNNs. The obtained text embeddings serve as the Key and Value matrices, which are fused with the Query matrix through a multi-head attention mechanism to yield feature representations incorporating both semantic information and graph structural information.
        
	\subsection{Design Space Exploration}
        To fully harness the semantic capabilities of LLMs for intelligent design space exploration, we propose the LLM4DSE methodology. This approach consists of three core steps: leveraging LLMs as optimizers to initialize or generate new solutions, evaluating these solutions, and updating both the optimal solutions and prompts accordingly. As shown in Fig. \ref{framework}, the iterative process commences with a task-specific prompting method PEODSE guiding LLMs to initialize solutions. These solutions are then processed by the Data Generator to produce graph and text embeddings, which are input into the Predictive Model for evaluation. The results of this evaluation update the optimal solution set and trigger the reconstruction of PEODSE, which in turn directs LLMs to generate new solutions. This cycle repeats until the iteration termination conditions (the maximum number of exploration design configurations) are met.
        
		{\bf{Why PEODSE?}} LLMs exhibit strong semantic comprehension capabilities, yet their performance on target tasks highly depends on carefully engineered prompts. While LLMMH \cite{xu2025} innovatively directs LLMs to perform solution generation for metaheuristics, the exponential growth in LLM parameters necessitates sophisticated prompt engineering to fully leverage domain-specific knowledge. The prompt design in LLMMH relied on minimal task descriptions, which limited the model’s depth of task understanding. While established techniques such as Zero-shot \cite{Radford2019}, Few-shot \cite{brown2020}, Optimization by PROmpting (OPRO) \cite{yang2024}, and Chain-of-Thought (CoT) \cite{wei2023} have shown promise in prompt engineering, we argue that enriching prompts with high-quality contextual information significantly enhances the quality of LLM-generated solutions. To this end, we propose PEODSE, a dedicated prompt formulation that provides comprehensive DSE task information, dynamically updated high-quality solutions, and reasoning traces for solution generation to reduce the probability of LLMs producing erroneous solutions. 
        
\begin{figure}[t]
			\centering
			\includegraphics[width=0.73\linewidth]{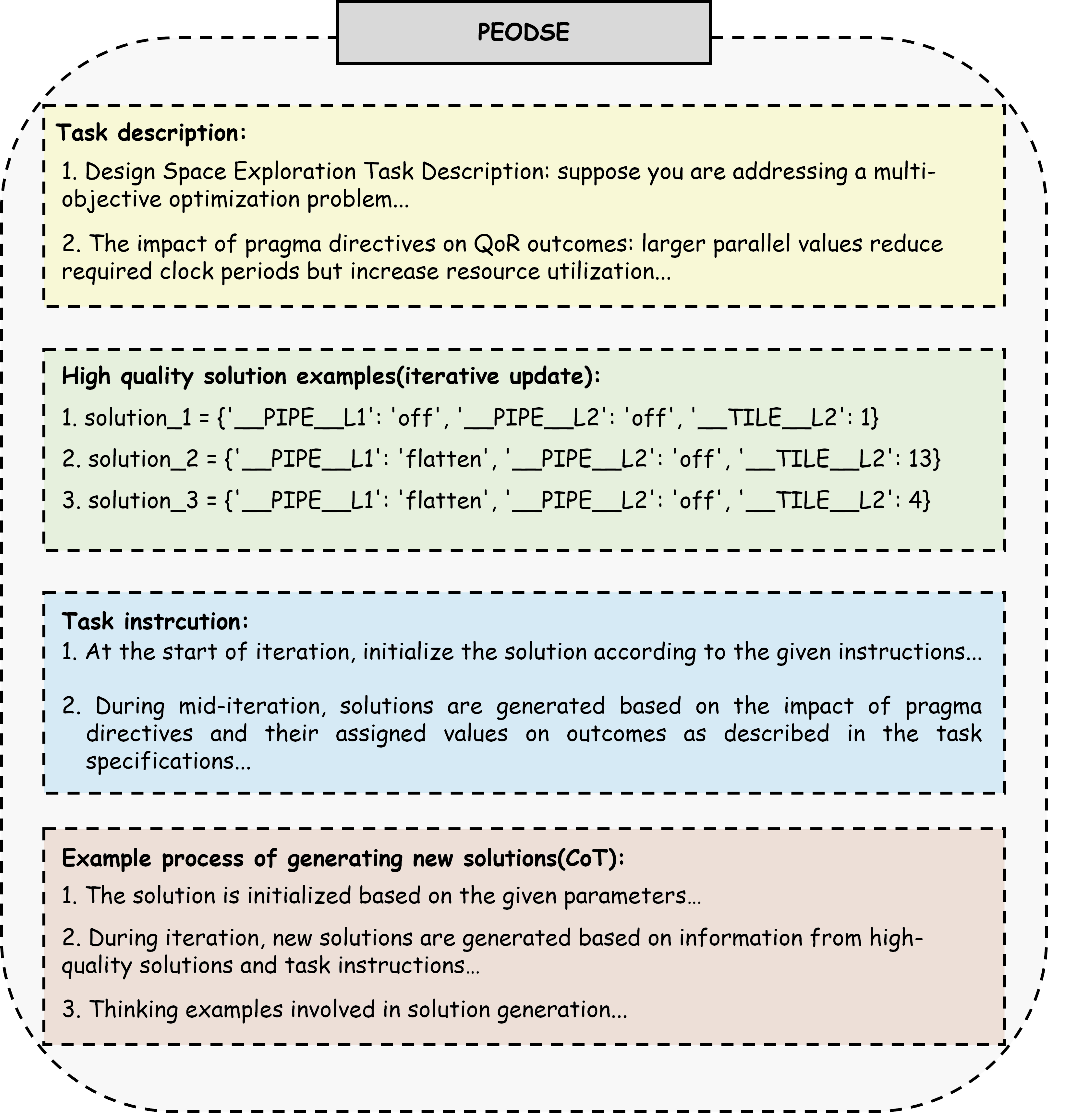}
			\caption{Illustration of the proposed prompting engineering strategy for DSE.}
			\label{PEODSE}
			\vspace{-5mm}
		\end{figure}

        As shown in Fig. \ref{PEODSE}, PEODSE comprises four components: task description, high-quality solution examples, task instruction, and solution generation exemplars. The task description introduces the task background and the impact of pragma directives with their values on QoR outcomes. High-quality solution examples, inspired by OPRO methodology, provide exemplary configurations and are dynamically updated during iterations. Task instruction delivers explicit guidance for LLMs to generate solutions. Within the task instruction, we incorporate the impact of pragma directives on QoR to enable more intelligent design space exploration (e.g., setting the pipeline directive to "off" reduces utilization such as LUTs while relatively increasing latency. Conversely, setting it to "flatten" yields the opposite effect.).  Solution generation exemplars incorporate CoT reasoning to demonstrate stepwise solution derivation processes. 
        
		\begin{table}[t]
			\setlength{\tabcolsep}{3pt}
			\centering
			\caption{The dataset in the training set consists of a total of 15 benchmarks, with 4,353 graph-text samples in total. Dataset Size denotes the number of graph-text samples employed per benchmark for training or inference purposes.}
			\begin{tabular}{c|c|c}
				\toprule
				\bf{Kernel} & \bf{Description} & \bf{Dataset Size} \\
				\midrule
				\textit{adi} &\scriptsize{Alternating direction implicit solver} & 322\\
				\textit{aes} & \scriptsize{A common block cipher} & 45 \\
				\textit{atax} &\scriptsize{Matrix transpose and vector multiplication} & 227\\
				\textit{bicg} &\scriptsize{BiCG sub kernel of bicgstab linear solver} & 347\\
				\textit{doitgen} &\scriptsize{multi-resolution analysis kernel} & 74\\
				\textit{fdtd-2d} &\scriptsize{2-D finite different time domain kernel} & 289\\
				\textit{gemm-blocked} &	\scriptsize{A blocked version of matrix multiplication} & 243 \\
				\textit{gemm-ncubed} & \scriptsize{Dense matrix multiplication} & 362\\
				\textit{gemver} &\scriptsize{Vector multiplication and matrix addition} & 464\\
				\textit{gemm-p} &\scriptsize{Matrix-multiply} & 303\\
				\textit{gesummv} &\scriptsize{Scalar, vector and matrix multiplication} & 159\\
				\textit{mvt} &\scriptsize{matrix vector product and transpose} & 476\\
				\textit{spmv-crs} & \scriptsize{Sparse matrix-vector multiplication(variable-length)}& 73\\
				\textit{spmv-ellpack} &\scriptsize{Sparse matrix-vector multiplication(fixed-size)} & 114\\
				\textit{2mm} &\scriptsize{2 matrix multiplications}& 388\\
				\textit{3mm} &\scriptsize{3 matrix multiplications} & 467\\ 
				\bottomrule
			\end{tabular}
			\vspace{-4mm}
			\label{ts}
		\end{table}
	
    \section{Experiments}
    \subsection{Experimental Setup}
	Based on the dataset from GNN-DSE \cite{Sohrabizadeh2022}, as detailed in Section \ref{III-A}, we construct a multimodal graph learning dataset for QoR prediction, implementing a more distinct partitioning scheme than GNN-DSE. As depicted in Table. \ref{ts}, we select five target kernels from MachSuite \cite{Reagen2014} and ten target kernels from Polyhedral benchmarks \cite{Yuki2010} as our training set. 70\% of the training set is allocated for training, 15\% for testing, and 15\% for validation. All GNN models employ a hidden layer dimension of 128, optimized via Adam with a batch size of 64, learning rate of 0.001, and 500+ iterations. Experiments execute on the AMD Ultrascale+ MPSoC ZCU104 platform. Benchmarks synthesize using Vitis-HLS 2022.1 and Vivado 2022.1 to collect ground-truth Latency and resource utilization (LUT, DSP, FF, BRAM) metrics, which serve as training labels.
	
    \subsection{Evaluation of HLS QoR Prediction Accuracy}
    We first evaluate HLS QoR prediction performance of the proposed MPM. Model training employs the dataset detailed in Table \ref{ts}. To demonstrate generalization capability, inference is performed on completely unseen kernels, with dataset specifics presented in Table \ref{is}. Our comparative analysis includes state-of-the-art (SOTA) approaches: GNN-DSE \cite{Sohrabizadeh2022}, HGBO \cite{Kuang2023}, IronMan-Pro \cite{Wu2023}, PROGSG\cite{Qin2024}. Experimental results demonstrate that MPM achieves the lowest prediction errors across all targets: 0.3870 (Latency), 0.0004 (LUT), 0.0004 (DSP), 0.0015 (FF), 0.0005 (BRAM). Compared with ProgSG, our approach significantly outperforms it by up to 10.25$\times$. MPM's superior performance originates from its enhanced multimodal feature fusion capability and utilization of text feature extractors with stronger semantic comprehension, which effectively establish correspondences between textual features and source code characteristics. 
    
    We also performed ablation study by comparing GNN-only and LM-only with MPM. The results reveal that standalone LM outperform ECoGNN, highlighting the untapped value of code semantics for QoR prediction. By fusing graph features from CDFGs (via ECoGNN) with textual features (via LM) under the multimodal graph learning framework, MPM significantly outperforms GNN-only approaches. 
	
	\begin{table}[t]
		\setlength{\tabcolsep}{1.5pt}
		\centering
		\caption{Kernels used for inference and design space exploration.}
		\begin{tabular}{c|c|c|c}
			\toprule
			\bf{Kernel} & \bf{Description} & \bf{ Dataset Size} & \bf{\# Design configs}\\
			\midrule
			\textit{heat-3d} & \scriptsize{Heat equation over 3D data domain} & 225 & 71511\\
			\textit{jacobi-1d} & \scriptsize{1-D Jacobi stencil computation}& 222 & 2871\\
			\textit{jacobi-2d} &\scriptsize{2-D Jacobi stencil computation} & 524 & 7609187\\
			\textit{nw} &	\scriptsize{A dynamic programming algorithm} & 386 & 6615\\
			\textit{seidel-2d} &\scriptsize{2-D Seidel stencil computation}& 56 & 10919\\
			\textit{stencil} & \scriptsize{A two-dimensional stencil computation} & 524 & 7591\\
			\bottomrule
		\end{tabular}
		\label{is}
        \vspace{-2mm}
	\end{table}

    \begin{figure}[t]
		\centering		
		\includegraphics[width=0.87\linewidth]{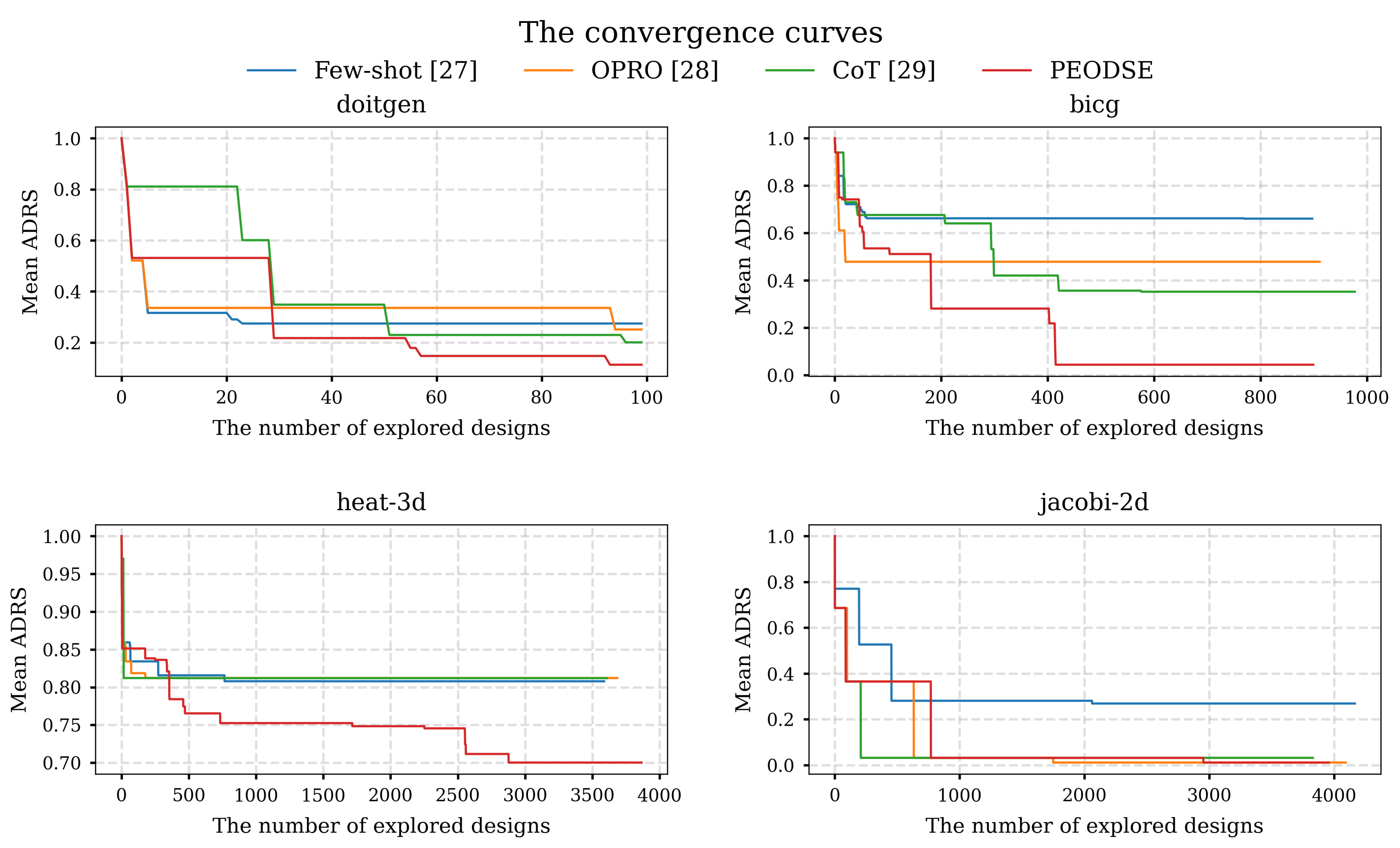}
		\caption{Convergence curves achieved by different prompting strategies.}
		\label{pgr}
		\vspace{-4mm}
	\end{figure}
    
	\begin{table}[t]
		\setlength{\tabcolsep}{3pt}
        \footnotesize
		\centering
		\caption{RMSE loss of MPM and SOTA models on unseen applications.}
		\begin{tabular}{l|c|c|c|c|c|c}
			\toprule
			{\bf{Model}}&{\bf{Latency}}&{\bf{LUT}}&{\bf{DSP}}&{\bf{FF}}&{\bf{BRAM}}&{\bf{All}} 
			\\
			\midrule
			{GNN-DSE\cite{Sohrabizadeh2022}}&0.7759&0.0025&0.0023&0.0078&0.0023&0.7909\\
			\hline
			{HGBO\cite{Kuang2023}}&0.5754&0.0127&0.0041&0.0064&0.0069&0.6082\\
			\hline
			{IronMan-Pro\cite{Wu2023}}&0.4201&0.0012&0.0015&0.0016&0.0007&0.4251\\
			\hline
			{PROGSG\cite{Qin2024}}&0.4061&0.0041&0.0018&0.0081&0.0025&0.4226\\
			\hline
			{ECoGNN-only}&0.4019&0.0016&0.0015&0.0016&0.0015&0.4084\\
			\hline
			{LM-only}&0.3920&0.0011&0.0011&0.0015&0.0009&0.3965\\
			\hline
			{MPM}&{\bf{0.3870}}&\bf{0.0004}&\bf{0.0004}&\bf{0.0015}&\bf{0.0005}&\bf{0.3898}\\
			\bottomrule
		\end{tabular}
		\label{QoR}
		\vspace{-3mm}
	\end{table}
	
	\subsection{Evaluation of Design Space Exploration}
	To demonstrate the efficacy of our proposed PEODSE method over alternative prompting approaches in DSE tasks, we evaluate various prompting strategies using the LLM4DSE framework. The quality of approximate Pareto-optimal sets is quantified using the Average Distance from the Referenced Set (ADRS) metric \cite{Schafer2019}:	
	\begin{equation}
		ADRS(\Gamma,\Omega) = \cfrac{1}{|\Gamma|}\ \underset{\lambda \in \Gamma}{\sum}\ \underset{\mu \in \Omega}{\min}f\left(\lambda, \mu \right)
		\label{9}
	\end{equation}
	where $\Gamma$ represents the reference Pareto-optimal set, $\Omega$ denotes the approximate Pareto-optimal set, and the function $f$ computes the distance between $\lambda$ and $\mu$. A lower ADRS value indicates a higher accuracy of the approximate Pareto-optimal set relative to the reference. We employ GPT-3.5-turbo \cite{ouyang2022} as the LLM for comparative experiments on prompting strategies, utilizing OpenAI's API \footnote{https://openai.com/index/openai-api/}. The convergence curves achieved by different prompting methods are presented in Fig. \ref{pgr}. The results demonstrate that PEODSE outperforms other prompting methods. This advantage arises from PEODSE’s ability to integrate the strengths of existing methods, coupled with the domain-specific knowledge of how pragma directives impact QoR. As a result, LLM gain a deeper understanding of DSE tasks and consistently find high-quality design configurations.

    Using our proposed MPM-LLM4DSE framework with PEODSE prompting, we perform comparative experiments comparing it against several established methods:  multi-objective genetic algorithm (NSGA-II) \cite{Schafer2017}, simulated annealing (SA) \cite{Schafer2009}, ant colony optimization (ACO) \cite{Schafer2016}, and LLMMH \cite{xu2025}. The eThe experiments target kernels that are unseen for the training presented in Table \ref{is}, leveraging high performance LLM (GPT-4o \cite{gpt4o2024}, Qwen3-235B-A22B-Thinking-2507 \cite{yang2025}, with results summarized in Table \ref{DR}. To ensure a fair comparison, all algorithms are set to explore identical number of designs. The experimental results demonstrate that both LLM4DSE combinations achieve superior exploration accuracy compared to traditional metaheuristics and LLMMH. LLM4DSE (Qwen3) yields the lowest ADRS value, showing average improvements of 46.07\% over conventional metaheuristic algorithms and 21.39\% over LLMMH. Additionally, LLM4DSE reduces runtime relative to LLMMH due to optimized prompt design that minimizes the latency of LLM reasoning. The current implementation of LLM4DSE is based on API calls. By localizing large language models, communication latency can be greatly reduced, which in turn may significantly shorten the runtime of LLM4DSE.
    
	To further evaluate the effectiveness of LLM4DSE, we compare it with the exact DSE algorithm of GNN-DSE \cite{Sohrabizadeh2022} on benchmarks with large design spaces containing up to 7 million design configurations. For each benchmark, we record LLM4DSE’s runtime, then execute GNN-DSE for the same duration and report its ADRS. Table \ref{CA} shows that under identical time constraints, LLM4DSE identifies higher-quality design configurations, exhibiting a 32.66\% reduction in ADRS over GNN-DSE, demonstrating its superior performance in large design spaces.
	
	\begin{table}[t]
		\centering
		\setlength{\tabcolsep}{1pt}
		\tiny
		\caption{ADRS results and runtime on unseen applications.}
		\begin{tabular}{c | c | c | c | c | c | c}
			\toprule
			\bf{Benchmark}&\bf{NSGA-II \cite{Schafer2017}}&\bf{SA \cite{Schafer2009}}&\bf{ACO \cite{Schafer2016}}&\bf{LLMMH \cite{xu2025}}&\bf{LLM4DSE(GPT-4o)}&\bf{LLM4DSE(Qwen3)} \\
			\midrule
			heat-3d & 0.0756 & 0.0672 & 0.0784 & 0.0645 & 0.0602 & \bf{0.0486} \\
			jacobi-1d & 0.0320 & 0.0253 & 0.0255 & 0.0298 & \bf{0.0212} & 0.0228\\
			jacobi-2d & 0.0111 & \bf{0.0014} & 0.0201 & 0.0029 & 0.0086 & 0.0035\\
			seidel-2d & 0.0517 & 0.0551 & 0.0606 & 0.0492 & \bf{0.0391} & 0.0406\\\
			stencil& 0.0946 & 0.0742 & 0.0893 & 0.0742 & \bf{0.0636} & 0.0666 \\
                nw & 0.0028 & 0.1197 & 0.1812 & 0.0121 & 0.0018 & \bf{0.0010} \\
                \hline
			Average ADRS & 0.0447 & 0.0572 & 0.0758 & 0.0388 & 0.0324 & \bf{0.0305}\\
                \hline
			Overall Runtime (s) & 1081& 664 & 1216& 10941 & 7222 & 9847 \\
			\hline
			\multicolumn{4}{c|}{\bf{LLM4DSE(Qwen3) Improv. over NSGA-II}}&\multicolumn{3}{c}{\bf{31.77\%}}\\
			\multicolumn{4}{c|}{\bf{LLM4DSE(Qwen3) Improv. over SA}}&\multicolumn{3}{c}{\bf{46.68\%}}\\
			\multicolumn{4}{c|}{\bf{LLM4DSE(Qwen3) Improv. over ACO}}&\multicolumn{3}{c}{\bf{59.76\%}}\\
			\multicolumn{4}{c|}{\bf{LLM4DSE(Qwen3) Improv. over LLMMH}}&\multicolumn{3}{c}{\bf{21.39\%}}\\
			\bottomrule
		\end{tabular}
		\label{DR}
		\vspace{-3mm}
	\end{table}
	
	\begin{table}[t]
		\centering
		\renewcommand{\arraystretch}{1.2}
		\setlength{\tabcolsep}{1pt}
            \scriptsize
		\caption{Comparative analysis of ADRS values between GNN-DSE and LLM4DSE on large benchmark under identical time constraints.}
		\begin{tabular}{c | c | c | c | c | c }
			\toprule
			\bf{Benchmark}& \bf{\# Design configs}&\bf{Time(s)}&\bf{LLM4DSE(Qwen3)}&\bf{GNN-DSE\cite{Sohrabizadeh2022}}&\bf{Improv.(\%)}\\
			\midrule
			heat-3d&71511&3068&\bf0.0486&0.0689&30.78\\
			jacobi-2d&7609187&4462&\bf0.0035&0.0074&52.70\\
			seidel-2d&10919&581&\bf0.0666&0.0779&14.51\\
			\bottomrule
		\end{tabular}
		\label{CA}
		\vspace{-5mm}
	\end{table}

\section{Conclusion}
	This paper presents a novel LM-GNN framework for HLS QoR prediction, offering an efficient and accurate alternative to conventional HLS tools. Our analysis reveals that LMs outperforms GNN architectures in QoR prediction. Furthermore, the proposed PEODSE prompting methodology significantly enhances LLM4DSE's task comprehension capabilities, leading to improved DSE outcomes. This work establishes a new paradigm for HLS DSE, highlighting the untapped potential of LMs and LLMs in hardware design. Future work will focus on employing smaller, fine-tuned and task-specific models for local execution to alleviate the computational burden and exploring its application in cross-platform synthesis.
	
\bibliographystyle{IEEEtran}
\bibliography{ref}

@inproceedings{Wu2021,
	author = {Wu, Nan and others},
	title = {IronMan: GNN-assisted Design Space Exploration in High-Level
	Synthesis via Reinforcement Learning
	},
	year = {2021},
	isbn = {9781450383936},
	publisher = {ACM},
	booktitle = {GLSVLSI},
	pages = {39–44},
	numpages = {6},
}

@INPROCEEDINGS{Lin2022,
	author={Lin, Zhe and others},
	booktitle={DATE}, 
	title={PowerGear: Early-Stage Power Estimation in FPGA HLS via Heterogeneous Edge-Centric GNNs}, 
	year={2022},
	pages={1341-1346},
}

@inproceedings{Sohrabizadeh2022,
	author = {Sohrabizadeh, Atefeh and others},
	title = {Automated accelerator optimization aided by graph neural networks},
	year = {2022},
	publisher = {ACM},
	booktitle = {DAC},
	pages = {55–60},
	}

@article{Ferretti2022,
	author = {Ferretti, Lorenzo and others},
	title = {Graph Neural Networks for High-Level Synthesis Design Space Exploration},
	year = {2022},
	volume = {28},
	number = {2},
	issn = {1084-4309},
	journal = {ACM Trans. Des. Autom. Electron. Syst.},
	articleno = {25},
	numpages = {20},
}

@INPROCEEDINGS{Kuang2023,
	author={Kuang, Huizhen and others},
	booktitle={ICFPT}, 
	title={HGBO-DSE: Hierarchical GNN and Bayesian Optimization based HLS Design Space Exploration}, 
	year={2023},
	pages={106-114},
}

@INPROCEEDINGS{Gao2024,
	author={Gao, Mingzhe and others},
	booktitle={DATE}, 
	title={Hierarchical Source-to-Post-Route QoR Prediction in High-Level Synthesis with GNNs}, 
	year={2024},
	pages={1-6},
}

@ARTICLE{Wu2023,
	author={Wu, Nan and others},
	journal={IEEE Transactions on Computer-Aided Design of Integrated Circuits and Systems}, 
	title={IronMan-Pro: Multiobjective Design Space Exploration in HLS via Reinforcement Learning and Graph Neural Network-Based Modeling}, 
	year={2023},
	volume={42},
	number={3},
	pages={900-913}}

@article{xu2025,
	title={Intelligent4DSE: Optimizing High-Level Synthesis Design Space Exploration with Graph Neural Networks and Large Language Models}, 
	author={Lei Xu and others},
	journal={arXiv preprint arXiv:2504.19649},
	year={2025},
}

@inproceedings{Wang2020,
	author = {Wang, Zi and others},
	title = {Machine learning to set meta-heuristic specific parameters for high-level synthesis design space exploration},
	year = {2020},
	publisher = {IEEE},
	booktitle = {DAC},
	articleno = {93},
	numpages = {6},
}

@article{yao2025,
	title = {Decomposition based estimation of distribution algorithm for high-level synthesis design space exploration},
	journal = {Integration},
	volume = {100},
	pages = {102292},
	year = {2025},
	author = {Yuan Yao and others},

}

@article{Schafer2019,
	title={High-level synthesis design space exploration: Past, present, and future},
	author={Schafer, Benjamin Carrion and others},
	journal={IEEE Transactions on Computer-Aided Design of Integrated Circuits and Systems},
	volume={39},
	number={10},
	pages={2628--2639},
	year={2019},
	publisher={IEEE}
}

@inproceedings{Lattner,
	title={LLVM: A compilation framework for lifelong program analysis \& transformation},
	author={Lattner, Chris and others},
	booktitle={International symposium on code generation and optimization},
	pages={75--86},
	year={2004},
	organization={IEEE}
}

@inproceedings{Chris,
	title={Programl: A graph-based program representation for data flow analysis and compiler optimizations},
	author={Cummins, Chris and others},
	booktitle={ICML},
	pages={2244--2253},
	year={2021},
}

@article{
    tang2025,
    title={Understanding {LLM} Embeddings for Regression},
    author={Eric Tang and others},
    journal={Transactions on Machine Learning Research},
    year={2025},
    }

@article{Joshi2025,
	title={Transformers are Graph Neural Networks}, 
	author={Chaitanya K. Joshi},
	year={2025},
	journal={arXiv preprint arXiv:2506.22084},
	eprint={2506.22084},
}

@inproceedings{Devlin2019,
    title = "{BERT}: Pre-training of Deep Bidirectional Transformers for Language Understanding",
    author={Jacob Devlin and others},
    booktitle = "Proceedings of the 2019 Conference of the North {A}merican Chapter of the Association for Computational Linguistics",
    year = "2019",
    pages = "4171--4186",
}

@inproceedings{clark2019,
	title = {What Does BERT Look at? An Analysis of {BERT}{'}s Attention},
	author = {Clark, Kevin  and others},
	booktitle = {ACL Workshop BlackboxNLP},
	year = {2019},
	publisher = {ACL},
	pages = {276--286},
}

@inproceedings{Zhou2023,
	title = {{C}ode{BERTS}core: Evaluating Code Generation with Pretrained Models of Code},
	author = {Zhou, Shuyan  and others},
	booktitle = {EMNLP},
	year = {2023},
	publisher = {ACL},
	pages = {13921--13937},
}

@inproceedings{Finkelshtein2024,
	author = {Finkelshtein, Ben and others},
	title = {Cooperative graph neural networks},
	year = {2024},
	booktitle = {ICML},
	articleno = {546},
	numpages = {27},
}

@article{Ba2016,
	title={Layer Normalization}, 
	author={Jimmy Lei Ba and others},
	year={2016},
	journal={arXiv preprint arXiv:1607.06450}
}

@inproceedings{Jang2017,
    title={Categorical Reparameterization with Gumbel-Softmax},
    author={Eric Jang and others},
    booktitle={ICLR},
    year={2017},
}

@inproceedings{bao2022,
 author = {Bao, Hangbo and others},
 booktitle = {NeurIPS},
 pages = {32897--32912},
 title = {VLMo: Unified Vision-Language Pre-Training with Mixture-of-Modality-Experts},
 volume = {35},
 year = {2022}
}

@book{Hamilton,
	title={Graph representation learning},
	author={Hamilton, William L},
	year={2020},
	publisher={Morgan \& Claypool Publishers}
}

@inproceedings{Li2016,
    author = {Li, Yujia and others},
    title = {Gated Graph Sequence Neural Networks},
    booktitle = {ICIR},
    year = {2016},
}

@inproceedings{vaswani2023,
	author = {Vaswani, Ashish and others},
	title = {Attention is all you need},
	year = {2017},
	booktitle = {NeurIPS},
	pages = {6000–6010},
}

@inproceedings{Qin2024,
	author = {Qin, Zongyue and others},
	title = {Cross-Modality Program Representation Learning for Electronic Design Automation with High-Level Synthesis},
	year = {2024},
	publisher = {ACM},
	booktitle = {MLCAD},
}

@inproceedings{Radford2019,
	title={Language Models are Unsupervised Multitask Learners},
	author={Alec Radford and others},
	year={2019},
	url={https://api.semanticscholar.org/CorpusID:160025533},
}

@inproceedings{brown2020,
	author = {Brown, Tom B. and others},
	title = {Language models are few-shot learners},
	year = {2020},
	booktitle = {NeurIPS},
}

@inproceedings{
    yang2024,
    title={Large Language Models as Optimizers},
    author={Yang, Chengrun and others},
    booktitle={The Twelfth International Conference on Learning Representations},
    year={2024}
}

@inproceedings{wei2023,
	author = {Wei, Jason and others},
	title = {Chain-of-thought prompting elicits reasoning in large language models},
	year = {2022},
	booktitle = {NeurIPS},
}

@inproceedings{Reagen2014,
	title={Machsuite: Benchmarks for accelerator design and customized architectures},
	author={Reagen, Brandon and others},
	booktitle={IISWC},
	pages={110--119},
	year={2014},
	organization={IEEE},
}

@misc{Yuki2010,
	author= {T. Yuki and others},
	year  = {2010},
	title = {PolyBenchC, {4.2.1}},
	note  = {\url{https://github.com/MatthiasJReisinger/PolyBenchC-4.2.1}Accessed on 2025-3-5}
}

@inproceedings{ouyang2022,
	author = {Ouyang, Long and others},
	title = {Training language models to follow instructions with human feedback},
	year = {2022},
	booktitle = {NeurIPS},
}

@article{Schafer2017,
	author = {Schafer, Benjamin Carrion},
	title = {Parallel High-Level Synthesis Design Space Exploration for Behavioral IPs of Exact Latencies},
	year = {2017},
	publisher = {ACM},
	volume = {22},
	number = {4},
	journal = {ACM Trans. Des. Autom. Electron. Syst.},
}

@INPROCEEDINGS{Schafer2009,
	author={Schafer, Benjamin Carrion and others},
	booktitle={International Symposium on VLSI Design, Automation and Test}, 
	title={Adaptive Simulated Annealer for high level synthesis design space exploration}, 
	year={2009},
	pages={106-109},
}

@ARTICLE{Schafer2016,
	author={Carrion Schafer, Benjamin},
	journal={IEEE Transactions on Computer-Aided Design of Integrated Circuits and Systems}, 
	title={Probabilistic Multiknob High-Level Synthesis Design Space Exploration Acceleration}, 
	year={2016},
	volume={35},
	number={3},
	pages={394-406},
}

@article{gpt4o2024,
	title={GPT-4o System Card}, 
	author={OpenAI and others},
	year={2024},
	journal={arXiv preprint arXiv:2410.21276}, 
}

@article{yang2025,
	title={Qwen3 Technical Report}, 
	author={An Yang and others},
	year={2025},
	journal={arXiv preprint arXiv:2505.09388}, 
}

\end{document}